\documentclass{ws-ijmpd}

\usepackage[super,sort&compress,merge,comma]{natbib}

\usepackage{amsfonts}
\usepackage{amsmath}
\usepackage{amssymb}
\usepackage{mathrsfs}
\usepackage{mathtools}
\usepackage[colorlinks]{hyperref}
\usepackage[hyphenbreaks]{breakurl}
\usepackage{etoolbox}
\usepackage{graphicx}
\usepackage{color}%
\usepackage{xcolor}

\definecolor{darkmidnightblue}{rgb}{0.0, 0.2, 0.4}
\definecolor{darkpowderblue}{rgb}{0.0, 0.2, 0.6}
\definecolor{darkslateblue}{rgb}{0.28, 0.24, 0.55}
\definecolor{magentanew}{rgb}{0.4, 0., 0.4}
\hypersetup{breaklinks=true,colorlinks=true,linkcolor=darkmidnightblue,citecolor=darkmidnightblue,filecolor=darkmidnightblue,urlcolor=darkmidnightblue}
\makeatletter
\patchcmd{\NAT@citex}
  {\@citea\NAT@hyper@{
\NAT@nmfmt{\NAT@nm}
\hyper@natlinkbreak{\NAT@aysep\NAT@spacechar}{\@citeb\@extra@b@citeb}
\NAT@date}}
  {\@citea\NAT@nmfmt{\NAT@nm}
\NAT@aysep\NAT@spacechar\NAT@hyper@{\NAT@date}}{}{}
\patchcmd{\NAT@citex}
  {\@citea\NAT@hyper@{
\NAT@nmfmt{\NAT@nm}
\hyper@natlinkbreak{\NAT@spacechar\NAT@@open\if*#1*\else#1\NAT@spacechar\fi}
{\@citeb\@extra@b@citeb}
\NAT@date}}
  {\@citea\NAT@nmfmt{\NAT@nm}
\NAT@spacechar\NAT@@open\if*#1*\else#1\NAT@spacechar\fi\NAT@hyper@{\NAT@date}}
  {}{}

\def\apj{Astrophys.~J.}
\def\apjl{Astrophys.~J.~Lett.}

\def\prd{Phys.~Rev.~D}
\def\jhep{JHEP}

\def\mpla{Mod.~Phys.~Lett.~A}
\def\ijmpa{Int.~J.~Mod.~Phys.~A}

\def\atmp{Adv.~Theor.~Math.~Phys.}
\def\cqgra{Class.~Quant.~Grav. }
\def\gregr{Gen.~Rel.~Grav.}

\def\nuphb{Nucl.~Phys.~B}

\def\phlb{Phys.~Lett.~B}
\def\anphy{Annals~Phys.}

\def\jmp{J.~Math.~Phys.}

\def\prl{Phys.~Rev.~Lett.}
\def\rvmp{Rev.~Mod.~Phys.}

\def\lrr{Living~Rev.~Rel.}

\def\adtmp{Adv.~Theor.~Math.~Phys.}
\def\ijtp{Int.~J.~Theor.~Phys.}

\def\ijmpd{Int.~J.~Mod.~Phys.~D}
\def\cejp{Central~Eur.~J.~Phys.}

\newcommand{\ssymbol}[1]{{\@fnsymbol{#1}}}

\begin{document}

\markboth{A. Danehkar et al.}
{Dual Fields of Massive/Massless Gravitons in IR/UV Completions}

\catchline{}{}{}{}{}
%
%%%%%%%%%%%%%%%%%%%%%%%%%%%%%%%%%%%%%%%%%%%%%%%%%%%%%%%%%%%%%%%%%%%%

\title{Dual Fields of Massive/Massless Gravitons\\ in IR/UV Completions\footnote{This essay received an Honorable Mention in the 2021 Essay Competition of the Gravity Research Foundation.}  }

\author{Ashkbiz Danehkar$^{\ssymbol{2},\ssymbol{6}}$, Hassan Alshal$^{\ssymbol{3},\ssymbol{4},\ssymbol{7}}$, and Thomas L. Curtright$^{\ssymbol{5},\ssymbol{8}}$}

\address{$^{\ssymbol{2}}$Department of Astronomy, University of Michigan,\\ Ann Arbor, Michigan 48109, USA\\\smallskip
$^{\ssymbol{3}}$Department of Chemistry and Physics,~Lincoln University,\\ Pennsylvania 19352, USA\\\smallskip
$^{\ssymbol{4}}$Department of Physics, Cairo University,\\ Giza 12613, Egypt\\\smallskip
$^{\ssymbol{5}}$Department of Physics, University of Miami,\\ Coral Gables, Florida 33146, USA\\\smallskip
$^{\ssymbol{6}}$danehkar@umich.edu\\
$^{\ssymbol{7}}$halshal@sci.cu.edu.eg\\
$^{\ssymbol{8}}$curtright@miami.edu}

\maketitle

\begin{history}
\received{21 May 2021}
%\revised{Day Month Year}
\accepted{8 September 2021}
%\published{Day Month Year}
Published 28 October 2021
\end{history}

\begin{abstract}
In the holographic picture, the Brout-Englert-Higgs (BEH) mechanism in $d$-dimensional Yang-Mills theories is conjectured to provide a Higgs-like mechanism for gravity in $d+1$ dimensions, resulting in massive (or massless) gravitons in IR (or UV) completions. Accordingly, one could imagine dual (magnetic-type) fields of massive gravitons in the IR (low-energy) limit that are coupled to the curl of their own energy-momentum, as well as to the rotation of matter fields on large scales. This hypothesis, which might solve cosmological issues currently ascribed to dark matter and dark energy, needs to be examined by the future Laser Interferometer Space Antenna (LISA) mission using observations of gravitational waves emitted from extragalactic sources.
\end{abstract}

\ccode{PACS numbers: 04.50.$-$h, 95.30.Sf, 04.50.Kd}

%\tableofcontents
\bigskip
~\\~\\
According to the holographic principle,\citep{Hooft1993,*Susskind1995} a
theory of gravity could be related to a gauge field theory in one less
dimension. This leads to the AdS$_{d+1}$/CFT$_{d}$ correspondence,\citep{Maldacena1998,*Maldacena1999,*Witten1998} 
also known as gauge/gravity duality. The
holographic correspondence between $d$-dimensional Yang-Mills theories and
gravitational fields in $d+1$ dimensions thereby hints at a geometric counterpart of
the BEH mechanism.\citep{Porrati2002,*Porrati2003,*Aharony2006} 
In the context of the standard model of particle physics,
the electroweak $W$ and $Z$ bosons acquire mass through the BEH mechanism,\citep{Englert1964,*Higgs1964,*Guralnik1964} which is analogous to the
behavior of photons inside superconductors.\citep{Goldhaber2010,*Wilczek2012} 
While photons are massless in empty space, they are apparently massive inside
superconductors at very low temperatures, and hence described by the Maxwell--Proca
equations. 
From the holographic principle, 
a Higgs-like mechanism
analogue to that in Yang-Mills theories could be expected for gravitational
theories in one higher dimension. This could give a mass to gravitons in the IR
completion on a large scale, resulting in massive gravity (see Ref.\,\citenum{Hinterbichler2012,*deRham2014} for a general review of massive gravity in various circumstances).

%dual (magnetic-type) fields of massive gravitons

In the theory of general relativity, the graviton is taken to be massless, with
tidal and frame-dragging fields encoded in the Weyl conformal tensor
$C_{\mu\alpha\nu\beta}$.\citep{Maartens1998,*Ellis2009,*Danehkar2009} %in free space
The decomposed
parts of the Weyl tensor, $E_{\mu\nu}\equiv C_{\mu\alpha\nu\beta}u^{\alpha}u^{\beta
}$ and $H_{\mu\nu}\equiv -\frac{1}{2}\epsilon_{\mu\gamma\alpha\beta}C^{\alpha\beta
}{}_{\nu\lambda}u^{\gamma}u^{\lambda}$ (where $\epsilon_{\mu\gamma\alpha\beta}$ is the Levi-Civita symbol), are analogous to the electric and
magnetic fields in Maxwell theory, thereby suggesting gravitational
electric-magnetic duality (see Ref.\,\citenum{Danehkar2019} for a review of this idea). By analogy
with the four-potential vector $A_{\mu}$ in Maxwell theory (spin 1), it is standard lore to anticipate a
2nd-rank symmetric tensor $h_{\mu\nu}$ in gravitational theories (spin 2). According to
the electric-magnetic invariance, the electromagnetic potential vector $A_{\mu}$ is dual
to a $(d-3)$-form potential field $\tilde{A}_{\mu_{1}\cdots\mu_{d-3}}$ in
$d$ dimensions, while it is dual to itself in $d=4$. 

The extension of the
electric-magnetic duality principle to $d$-dimensional gravitational theories
implies that the symmetric tensor $h_{\mu\nu}$ of a massless graviton is dual to
a mixed symmetry $(d-3,1)$ tensor $\tilde{h}_{[\mu_{1}\cdots\mu_{d-3}]\nu}$ of
rank $d-2$ (also called Curtright fields),\citep{Curtright1985,Hull2000,*Hull2001,*West2001} 
whereas
the $h_{\mu\nu}$ of a massive graviton is dual to a mixed symmetry $(d-2,1)$ tensor
$\tilde{h}_{[\mu_{1}\cdots\mu_{d-2}]\nu}$ of rank $d-1$.\citep{Curtright1980,*Curtright2019,Curtright2019a,Alshal2019,*Alshal2020,Gonzalez2008} 
In $d=4$, the massless field $h_{\mu\nu}$ is dual to itself, whereas
the massive field $h_{\mu\nu}$ is dual to a mixed symmetry $(2,1)$ tensor
$\tilde{h}_{[\mu\nu]\lambda}$. The number of propagating helicity modes of
$h_{\mu\nu}$ and its dual field is $d(d-3)/2$ for massless gravitons and
$(d+1)(d-2)/2$ for massive gravitons. In $d=4$, 
a massless graviton propagates with 2 helicity modes, whereas a massive graviton
propagates with 5 helicity modes.

We know that the (gravito-)magnetic part of the Weyl tensor of massless gravitons can be geometrically interpreted as 
frame-dragging vortexes around spinning massive objects.\citep{Owen2011,*Nichols2011,*Nichols2012,*Danehkar2020} 
Similarly, by comparison with $\tilde{A}_{\mu_{1}\cdots\mu_{d-3}}$, 
one could speculate about magnetic-like effects introduced by $\tilde{h}_{[\mu_{1}\cdots\mu_{d-3}]\nu}$. 
This stems from the fact that the \textit{geometric} Lanczos tensor $H_{[\mu\nu]\alpha}$ generating the Weyl tensor $C_{\mu\nu\alpha\beta}$ \citep{Lanczos1962,*Takeno1964} has gauge symmetries similar to those of the Curtright fields, 
which can be used to predict frame-dragging effects for the Curtright fields 
in a way same as \textit{gravitomagnetism} in general relativity.

The Lagrangian density for dual fields of massless gravitons in $d$ dimensions can
be expressed as $\mathcal{L}_{{\mathrm{massless}}}=\mathcal{F}_{\mu}{}^{\nu
}\mathcal{F}_{\nu}{}^{\mu}$, where $\mathcal{F}_{\mu}{}^{\nu}$ is
the curl of dual fields defined as:\citep{Curtright1985,Boulanger2008}
\begin{equation}
\mathcal{F}_{\mu}{}^{\nu}\equiv\frac{1}{(d-3)!}\epsilon^{\lambda_{1}%
\cdots\lambda_{d-2}\nu}\partial_{\lambda_{1}}\tilde{h}_{[\lambda_{2}\cdots
\lambda_{d-2}]\mu}. \label{eq1}
\end{equation}
For dual fields of massive gravity, the Lagrangian density is as follows:\citep{Curtright1980,*Curtright2019,Curtright2019a,Alshal2019,*Alshal2020,Gonzalez2008}
\begin{align}
\mathcal{L}_{{\mathrm{massive}}}  =  \mathcal{K}_{\mu}{}^{\nu}\mathcal{K}%
_{\nu}{}^{\mu}+ & \frac{(-1)^{d}}{(d-2)!}m^{2}\left(  \tilde{h}_{[\lambda
_{1}\cdots\lambda_{d-2}]\mu}\tilde{h}^{[\lambda_{1}\cdots\lambda_{d-2}]\mu
}\right.  \nonumber\\
&  \left.  -(d-2)\tilde{h}_{[\lambda_{1}\cdots\lambda_{d-3}]\nu|}{}^{\nu
}\tilde{h}^{[\lambda_{1}\cdots\lambda_{d-3}]\rho|}{}_{\rho}\right)  , \label{eq2}
\end{align}
where $m$ is the graviton mass, and $\mathcal{K}_{\mu}{}^{\nu}$ is the curl of dual fields of massive
gravity:%
\begin{equation}
\mathcal{K}_{\mu}{}^{\nu}\equiv\frac{1}{(d-2)!}\epsilon^{\lambda_{1}%
\cdots\lambda_{d-1}\nu}\partial_{\lambda_{1}}\tilde{h}_{[\lambda_{2}\cdots
\lambda_{d-1}]\mu}. \label{eq3}
\end{equation}
The Lagrangian density (\ref{eq2}) is analogue to the Proca action for massive photons,
$\mathcal{L}_{{\mathrm{Proca}}}=-\frac{1}{4}F_{\mu\nu}F^{\mu\nu}+\frac{1}{2}m^{2}A_{\mu
}A^{\mu}$.

The dual fields of massive gravitons can be coupled to the curl of any conserved,
energy--momentum tensor:\citep{Curtright1980,Alshal2019,*Alshal2020}
\begin{align}
& \left(  \square+m^{2}\right)  \tilde{h}_{[\lambda
_{1}\cdots\lambda_{d-2}]\nu}=\kappa P_{\lambda_{1}\cdots\lambda_{d-2}\nu}{}^{\alpha\beta}{}^{\gamma}
 \partial_{\alpha} \left(  \mathcal{T}_{\beta\gamma}+\mathfrak{t}_{\beta\gamma}\right) , \label{eq5}
\\
& P_{\lambda_{1}\cdots\lambda_{d-2}\nu}{}^{\alpha\beta}{}^{\gamma} \equiv  (d-3)\epsilon_{\lambda_{1}
\cdots\lambda_{d-2}}{}^{\alpha\beta}\eta{}_{\nu}{}^{\gamma}
+(d-1)\epsilon_{(\lambda_{1}
\cdots\lambda_{d-2}}{}^{\alpha\beta}\eta{}_{\nu)}{}^{\gamma}, \label{eq6}
\end{align}
where $P_{\lambda_{1}\cdots\lambda_{d-2}\nu}{}^{\alpha\beta}{}^{\gamma}$ is a Young symmetrizer, 
$\eta_{ab}={\mathrm{diag}}(-1,+1,+1,+1)$ the metric tensor,  
$\square \equiv \partial_{\mu}\partial^{\mu}$ the d'Alembert operator,
$\kappa$ a gravitational constant in $d$-dimensional spacetime,
$\mathcal{T}_{\alpha\beta}$ the energy--momentum tensor of arbitrary matter fields,
and $\mathfrak{t}_{\alpha\beta}$ the energy--momentum tensor of dual fields
of massive gravitons (see Ref.\,\citenum{Alshal2019,*Alshal2020}). 
The field equation (\ref{eq5}) is analogous with $(\square + m^{2}) A_{\mu}= J_{\mu}$ in Maxwell--Proca theory.

From Eq.~(\ref{eq5}), it follows the on-shell equation (see Eq.~35 in Ref.\,\citenum{Curtright2019a}) for the dual field strength $\mathcal{K}_{\mu}{}^{\nu}$:\citep{Curtright2019a,Alshal2019,*Alshal2020} 
\begin{align}
& \left(  \square+m^{2}\right)  \mathcal{K}_{\mu}{}^{\nu} =
 \kappa  (1-d)\square \left(  \mathcal{T}_{\mu}{}^{\nu}+\mathfrak{t}_{\mu}{}^{\nu}\right) 
+ \kappa \left(  \mathcal{\delta}_{\mu}{}^{\nu} \square -\partial_{\mu}\partial^{\nu}\right) (\mathcal{T}+\mathfrak{t}) , \label{eq7}
\end{align}
where $\mathcal{T}\equiv \mathcal{T}_{\lambda}{}^{\lambda}$ and $\mathfrak{t}\equiv \mathfrak{t}_{\lambda}{}^{\lambda}$, and 
${\delta}_{\mu}{}^{\nu}\equiv \eta_{\mu\lambda} \eta^{\lambda\nu}$ is the Kronecker delta.
The field equation (\ref{eq7}) is similar to the conventional one in massive gravity, $( \square+m^{2}) h_{\mu\nu} = \kappa \mathcal{T}_{\mu\nu}$, apart from its constant with an opposite sign ($d>3$) and the manifestly conserved trace term.
As shown in Ref.\,\citenum{Curtright2019a}, Eq.~(\ref{eq7}) leads to the Ogievetsky-Polubarinov theory,\citep{Ogievetsky1965} implying that the Curtright field is the exact dual of the Ogievetsky-Polubarinov model for massive gravitons.

According to Eqs. (\ref{eq5}) and (\ref{eq7}), the curl of massive gravitons in $d$-dimensional spacetime is associated with dual (magnetic-type) fields described by a $(d - 1)$-rank tensor of Young symmetry type $(D - 2, 1)$, which are coupled to the curl of their own energy-momentum, as well as the rotation of matter fields on large scales. 
In this picture, dual fields of massive gravity that strongly emerge in the IR completion could have major implications for some phenomena on large scales. 
Moreover, as graviton is massless in the UV limit, the theory of general relativity still remains valid within the solar system and high energy domains. 

It is argued that duality breaks asymptotic flatness of the Schwarzschild solution, and consequently fills the empty space surrounding isolated celestial objects, because of the extra constant that appears in the $g_{tt}$, which is responsible for creating the Barriola-Vilenkin monopole solution.\citep{Dadhich2000,*Dadhich1999} 
In massive dual gravity, we have a constant, with the negative sign (see also Eq. \ref{eq7}), that plays the similar role in creating the dual graviton. The constant appears after expanding the exponent term in the static spherically symmetric solution of the $h_{00}$ in the interaction theory of massive dual gravity (see Eqs. 39--40 in Ref.\,\citenum{Curtright2019a}). Thus, if gravitoelectromagnetism prescribed by general relativity was a good candidate as a Machian theory, then the massive gravity with its dual field could be too.

% implications for cosmology

Gravitons that acquire a mass in the IR completion could potentially offer some solutions 
to the current issues in cosmology, namely flat rotation curves of \textit{disk} galaxies and accelerating expansion of the universe since about 7 billion years ago. 
Gravitational forces mediated by massive gravitons are weakened at distances larger than the graviton's Compton wavelength 
$\lambda_{g} = \hbar / (m_{g} c^2)$ according to $ \sim \frac{1}{r} \exp (-r/ \lambda_{g})$, by contrast, 
dual (magnetic-type) fields produced by the rotation of matter fields and massive gravitons turn to be stronger than gravitomagnetic fields associated with massless gravitons. 
Massive gravity yields dual fields associated with the curl of arbitrary conserved energy-momentum that gravitationally confine stars rotating with galactic disks of spiral galaxies. 
Moreover, dual fields of massive gravity may lead to the cosmic acceleration at late times if 
graviton is massless in the UV, but massive in the IR. 
This may naturally explain why the acceleration 
of the cosmic expansion occurs at late times on large scales when intergalactic space predominantly got much cooler.
In particular, dual fields of massive gravity on a large scale may contribute to some preferentially orientations in spiral galaxies in filaments, similar to those reported in the literature.\citep{Trujillo2006,*Lee2007,*Tempel2013,*Zhang2013}

%gravitational waves of massive gravity and observations

Gravitational waves (GW) in general relativity ($d=4$) propagate at the speed of light with two tensor polarization modes: $+$ and $\times$. In contrast, the speed of gravitational waves in massive gravity could be frequency-dependent, slightly below the speed of light. 
Apart from the frequency-dependent speed, massive gravity in $d=4$ also generates five polarization modes: 
two tensors, two vectors, and one scalar (see the review by Ref.\,\citenum{Ezquiaga2018}). If graviton remains massless in the UV regime,
 gravitational waves detected from nearby sources within our Galaxy should perfectly agree with general relativity. 
However, graviton may enter its massive mode as it propagates on intergalactic scales in the IR limit.
In this IR/UV scenario, that makes massive/massless gravitons, GW emitted from a binary merger in a nearby galaxy 
could potentially carry three additional polarization modes, with a slight time delay between the GW and its electromagnetic counterpart. 
The presence of additional GW polarization modes can be investigated using 3-arm space-borne GW detectors such as the Laser Interferometer Space Antenna (LISA; planned for 2034), whereas 2-arm GW detectors similar to the Laser Interferometer Gravitational-Wave Observatory (LIGO) are inadequate for this analysis. Future space-borne GW defectors will decide whether gravitons propagate in the massive mode in the IR regime.

% Higgs-like mechanism for gravity

Vector bosons in Yang-Mills theories can acquire a rest mass through the BEH mechanism by ingesting the Higgs scalar field (spin 0). However, the Higgs field is supposed to provide the BEH mechanism for vector bosons (spin 1). %, 
As argued by Ref.\,\citenum{tHooft2007}, 
the holographic correspondence maps a 4-dimensional Yang-Mills theory onto a 5-dimensional theory of gravity that could also be projected onto a massive graviton in 4 dimensions. In this scenario, four massless scalar fields can assemble a Higgs-like mechanism for gravity in $d=4$, 
by which a massive graviton and a massive scalar boson survive,\citep{tHooft2007,Kakushadze2008,*Chamseddine2010} while the other three scalar fields are absorbed into the massive graviton.
This geometric counterpart of the Higgs boson should be detectable in our future high-energy particle physics experiments if the nature also supports the Higgs-like mechanism for gravity.

%\section*{Acknowledgments}

%This section should come before the References. Dedications and funding
%information may also be included
%here.

%\section{References}

%\bibliographystyle{ws-ijmpd}
%\bibliographystyle{apsrev4-1-v2}
%\bibliography{references}

%

\end{document}